\documentclass[fleqn,twoside]{article}
\usepackage{espcrc2}

\usepackage{graphicx}
\usepackage[figuresright]{rotating}

\newcommand{\AmS}{{\protect\the\textfont2
  A\kern-.1667em\lower.5ex\hbox{M}\kern-.125emS}}

\input babarsym

\newcommand\vus {\ensuremath{V_{\rm us}}}
\newcommand\vub {\ensuremath{V_{\rm ub}}}
\newcommand\vcd {\ensuremath{V_{\rm cd}}}

\newcommand\vcb {\ensuremath{V_{\rm cb}}}
\newcommand\vtd {\ensuremath{V_{\rm td}}}

\newcommand\vtb {\ensuremath{V_{\rm tb}}}

\newcommand\tentothethirtysix{\ensuremath{10^{36} \, {\mathrm \cm^{-2} \rm s^{-1}}}}

\newcommand\etap {\ensuremath{\eta^\prime}}
\def\CPV{\ensuremath{CPV}}
\def\ifb{\ensuremath{\mathrm{fb^{-1}}}}
\def\iab{\ensuremath{\mathrm{ab^{-1}}}}
\def\D{\ensuremath{D}}
\def\ptrue{\ensuremath{f_L}}

\def\BBbar{{\ensuremath{B\overline{B}}}}

\newcommand{\su}    [1]   {{\rm{SU({#1})}}}
\newcommand{\Uns}  [1]   {{\ensuremath{\Upsilon({\rm{#1 S})}}}}

\begin{document}

\title{The Super Flavor Factory}

\author{A. J. Bevan\address[QMUL]{Physics Department, Queen Mary, University of London,
        E1 4NS, UK.}%
        \thanks{This work is supported by PPARC and the DOE under contract DE-AC02-76SF00515.}
}

\begin{abstract}
The main physics goals of a high luminosity \epem\ flavor factory are discussed, including the possibilities to perform
detailed studies of the CKM mechanism of quark mixing, and constrain virtual Higgs and  Non-Standard Model particle
contributions to the dynamics of rare $B_{u,d,s}$ decays.  The large samples of $D$ mesons and $\tau$ leptons produced
at a flavor factory will result in improved sensitivities on $D$ mixing and lepton flavor violation searches,
respectively. One can also test fundamental concepts such as lepton universality to much greater precision than
existing constraints and improve the precision on tests of \CPT\ from $\B$ meson decays.  Recent developments in
accelerator physics have demonstrated the feasibility to build an accelerator that can achieve luminosities of ${\cal
O}(\tentothethirtysix)$. \vspace{1pc}
\end{abstract}

\maketitle

\section{INTRODUCTION}

Recent developments in accelerator physics show that it is feasible to construct an \epem\ collider with a luminosity
of \tentothethirtysix, which is a factor of fifty increase relative to the current \B-factories~\cite{Oide,seeman}.
This paper discusses the physics potential of a Super Flavor Factory (SFF) associated with such a collider.  The
physics potential of a SFF comes from vast samples of $B_{u,d,s}$, \D\ mesons and $\tau$ leptons that can be produced,
in addition to the flexibility of operating at different center of mass energies ($\sqrt{s}$). One can study $\Uns{n}$
decays where $n=1,2,3,4,5$, and perform precision measurements of the ratio $R=\sigma(\epem\to
hadrons)/\sigma(\epem\to\mu^+\mu^-)$. Detailed reports have been compiled on the potential of a
SFF~\cite{sffreports1,sffreports2,sffreports3}.

The remainder of these proceedings discuss precision tests of the Cabibbo-Kobayashi-Maskawa (CKM) quark-mixing
matrix~\cite{ckm1,ckm2}, new physics constraints from loop-dominated processes, rare charmless \B\ decays, tests of the
combined symmetry of charge-conjugation, parity and time reversal (\CPT), charm and $\tau$ physics opportunities and
the physics potential from analysing data accumulated at the $\Uns{1S,2S,3S,5}$ resonances.

\section{PRECISION CKM METROLOGY}

Violation of the combined symmetry of charge-conjugation and parity (\CP) was first seen in the decay of neutral
kaons~\cite{christenson}. All \CP\ violation (\CPV) in the Standard Model (SM) is the result of a single complex phase
in the CKM quark-mixing matrix. It was shown some time ago that \CPV\ is a necessary but insufficient constraint in
order to generate a net baryon anti-baryon asymmetry in the universe~\cite{sakharov}.  The other requirements for this
asymmetry are a non-thermal equilibrium in the expansion of the universe and baryon number violation. In the ensuing
years there has been a tremendous amount of activity to elucidate the role of \CPV\ in the SM. All measurements of
\CPV\ produced by the \babar~\cite{babar_nim} and Belle~\cite{belle_nim} experiments are consistent with the CKM
description of \CPV, and insufficient to explain the matter-antimatter asymmetry of the universe.

The SM description of \CPV\ for $B_{u,d}$ decays is manifest in the form of a triangle in a complex plane.  This
triangle has three non-trivial angles, $\alpha$, $\beta$, and $\gamma$, and two non-trivial sides with magnitudes
$|\vus\vub^*|/|\vcd\vcb^*|$ and $|\vtd\vtb^*|/|\vcd\vcb^*|$, where $V_{ij}$ are elements of the $3\times 3$ CKM matrix.
The limiting factor in the determination of the sides of the triangle are the magnitudes of the CKM matrix elements
\Vub\ and \Vtd. The first step toward understanding any new physics weak phase or amplitude contribution to the flavor
sector is to precisely understand the SM contributions. To this end, one has to overconstrain the parameters describing
the unitarity triangle before embarking on a quest to find deviations from SM behavior.

The angle $\beta$ is determined from a time-dependent analysis of $b\to c\overline{c}s$ decays~\cite{babarsin2betaprd}.
This determination is from tree-dominated processes that are theoretically clean and provides a baseline to compare
against the results from measurements of $b\to s$ and $c\overline{c}d$ transitions. The current results from the
\B-factories on this parameter are Refs.~\cite{babar_sin2beta,belle_sin2beta}. As the precision of these measurements
increase it will become increasingly important to improve our understanding of possible SM
pollution~\cite{sin2beta_smpollutiona,sin2beta_smpollutionb,sin2beta_smpollutionc}.  Open charm decays such as $\B^0\to
J/\psi \piz$~\cite{babar_jpsipiz,belle_jpsipiz} can be used to estimate this SM pollution~\cite{sin2beta_smpollutionc}.
A SFF will be able to improve on existing measurements of this angle and provide competitive results in the LHC era.

The measurement of $\alpha$ is more complicated than that of $\beta$~\cite{bevan2006,bianchi}. The parameters of the
time-dependent analysis of $b\to u\overline{u}d$ transitions do not provide a clean measurement of $\alpha$, but
measure an effective parameter dependent on tree and loop (penguin) contributions to the overall weak phase.  The most
stringent method for extracting $\alpha$ currently comes from the study of $\B\to\rho\rho$
decays~\cite{babarrhoprhomprlr12,babarrhoprhomr15prelim,bellerhoprhom}. This result has an 8-fold ambiguity, with
degenerate solutions.  The degeneracy can be resolved using the result of a time-dependent Dalitz-plot analysis of
$\Bz\to\pi^+\pi^-\pi^0$ decays~\cite{snyderquinn,quinnsilva}. The precision on $\alpha$ measured using $B\to \pi\pi$
decays is limited by the ability to measure $\Bz\to \piz\piz$. A SFF will enable us to perform a precision measurement
of this mode. There will be sufficient statistics to measure time-dependent asymmetry parameters through $\piz\piz$
decays with photon conversions, and a precision study of $\Bz\to\piz\piz$ means that an isospin analysis of
$B\to\pi\pi$ decays will become an increasingly important contribution to the overall constraint on $\alpha$.  The
constraint on $\alpha$ obtained at a SFF will be a combination of results from all of these channels. The hadronic
environment of LHCb results in difficulties in studying channels with neutral particles in the final state and some
channels required to constrain penguins in  $b\to u\overline{u}d$ transitions will only be accessible to a SFF.

A precision measurement of $\gamma$ will require a systematic study of the many methods proposed in the
literature~\cite{all_gamma_at_beauty2006}. One of the most promising channels to extract $\gamma$ is $\B\to DK$, where
$D^0\to K^0_s \pi^+\pi^-$, and the structure of the $K^0_s\pi^+\pi^-$ Dalitz plot is used in the
fit~\cite{gamma_at_beauty2006}.  A SFF with 50\,\iab\ should be able to measure $\gamma$ at the level of $2^\circ$ with
this method~\cite{bondar}.  The precision of the constraint on $\gamma$ using the Atwood-Dunietz-Soni~\cite{ads} and
Gronau-London-Wyler~\cite{glw} methods is expected to be dominated by Dalitz-plot model uncertainties at a SFF.

\begin{figure*}[!ht]
\begin{center}
\includegraphics[width=110mm]{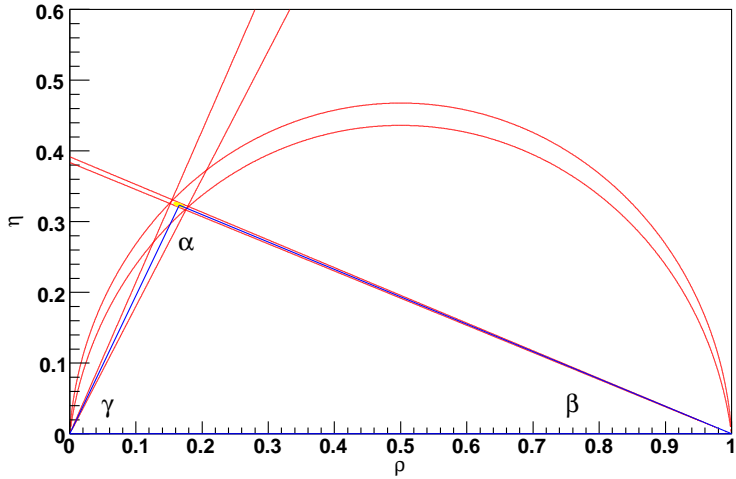}
  \caption{
  A prediction of the constraint on the unitarity triangle obtained
  at a SFF using a 50\,\iab\ data sample.  The contours represent 68.3\%
  confidence level intervals in the $\rho -\eta$ plane (the parameters which are
  defined in the Wolfenstein expansion of the CKM matrix~\cite{Wolfenstein}). }
  \label{fig:utprediction}
\end{center}
\end{figure*}

A SFF will be able to test the closure of the unitarity triangle to a few degrees with a data-set of 50\,\iab.  The
projected sensitivities for $\alpha$, $\beta$ and $\gamma$ are $2^\circ$, $0.2^\circ$ and $2^\circ$, respectively as
shown in Figure~\ref{fig:utprediction}.  This level of sensitivity is comparable to the expectations of an upgraded LHCb
experiment~\cite{lhcbupgrade}. In addition to performing the primary measurements of the unitarity triangle
angles, a SFF has the ability to perform measurements that will enable better determination of the SM theoretical
pollution to the angle measurements. This is a critical aspect of searching for manifestations of a NP weak phase.

\section{NEW PHYSICS FROM LOOPS}

The Higgs particle and supersymmetry are introduced to the SM as the standard way to elucidate the mass generation
mechanism~\cite{higgs} and resolve the hierarchy problem~\cite{weinberg,susy_primer}.  The energy scale of the Higgs
and NP contributions are widely expected to be introduced below $\sim 1\tev$.  When extending the SM to accommodate new
particles at this scale, one introduces couplings in the flavor sector that will impinge on low energy measurements of
flavour-changing neutral currents (FCNCs) and processes dominated by penguin amplitudes at a \B-factory. Most calculations with
a NP contribution to the effective Lagrangian introduce a fine-tuning problem by setting the NP flavor parameters to
zero. The rest of this section highlights a few specific examples of processes that can be used to constrain the
effects of NP.

\subsection{\boldmath Measurements of $\Delta S$}

The \B-factories have recently observed \CPV\ in $B\to \etap K^0$ decays~\cite{etapk_babar,etapk_belle}.  These $b \to
s$ penguin processes are probes of NP, and have the most precisely measured time-dependent \CP\ asymmetry parameters of
all of the penguin modes.  Any deviation $\Delta S$ of the measured asymmetry parameter $S_{\eta^\prime K^0}$ from
$\sin 2\beta$ is an indication of NP (for example, see Refs.~\cite{sffreports1,sffreports2}). The anticipated
precision on $\Delta S$ for $\eta^\prime K^0$ is $\sim 0.015$ with a data sample of 50\,\iab. In addition to relying on
theoretical calculations of the SM pollution to these
decays~\cite{etapk_smpollutiona,etapk_smpollutionb,etapk_smpollutionc}, it is possible to experimentally constrain the
SM pollution using \su{3} symmetry~\cite{etapksuthree}.  This requires precision knowledge of the branching fractions
of the $B$ meson decays to the following pseudo-scalar pseudo-scalar ($PP$) final states $\piz\piz, \piz\eta,
\piz\etap, \eta\eta, \etap\eta, \etap\etap$ ~\cite{etaeta_babar,etaeta_babartwo}.  As these $PP$ channels
have several neutral particles in each of their final states, they will be very difficult to study in a hadronic
collider environment. The decay $B\to\phi K^0_S$ is another theoretically clean $b \to s$ penguin channel to
search for signs of NP~\cite{sffreports1,sffreports2,sffreports3}.  The expected 50\,\iab\ SFF precision on
$\Delta S$ for $K^+K^-K^0$, which contains the mode $\phi K^0$, is $\sim 0.017$.

\subsection{\boldmath $b\to s \gamma$}

The $b \to s \gamma$ penguin decays provide one of the most sensitive constraints on possible NP (for example, see
Ref.~\cite{gambino}). The existing experimental measurements and interpretation of results has been widely debated (for
example, see Ref.~\cite{hurth}). Of interest are both inclusive and exclusive decays where the rate measurement is used
to constrain the mass of Higgs particles (see Refs.~\cite{gambino,hurth}). However in addition to this, it is possible
to measure time integrated and time dependent \CP\ asymmetry parameters.  These measurements also provide stringent
constraints on NP models, and the expected precision on the charge asymmetry with 10\,\iab\ at a SFF is $\sim 1\%$.

\subsection{\boldmath $b \to s ll$}

The $b \to s ll$ ($l=e, \mu$) processes are the result of FCNCs.  The
forward-backward asymmetry of these FCNC processes are sensitive to NP contributions.  Recently the \babar\ and Belle
experiments started to study the asymmetry~\cite{babarsll,bellesll} and a high statistics evaluation of the
forward-backward asymmetry is required to elucidate the nature of any NP contribution to these decays. The SFF will be
able to study both $e^+e^-$ and $\mu^+\mu^-$ final states. The physics reach of a SFF with these decays will be
competitive with any results from an upgraded LHCb experiment.  With 50\,\iab\ it is expected that the effective
parameters related to the Wilson coefficients $C_{9,10}$ can be measured to 10-15\% from the forward backward
asymmetry in $B\to K^* l^+l^-$.

\subsection{\boldmath $B\to VV$ decays}

The angular analysis of $B\to VV$ decays (where $V$ is a vector meson) provides eleven observables (six amplitudes and
five relative phases) that can be used to test theoretical calculations~\cite{Dunietz}. The hierarchy of $A_0$, $A_+$,
and $A_-$ amplitudes obtained from a helicity (or $A_0$, $A_\parallel$, and $A_\perp$ in the transversity basis)
analysis of such decays allows one to search for possible right handed currents in any NP contribution to the total
amplitude. For low statistics studies, a simplified angular analysis is performed where one measures the fraction of
longitudinally polarised events defined as $\ptrue=|A_0|^2/\sum{|A_i|^2}$, where $i=-1,0,+1$. Current data for penguin-dominated processes ($\phi K^*(892)$~\cite{babar_phikst,belle_phikst}, $K^*(892)\rho$~\cite{babar_rhokst,belle_rhokst})
that are observed to have non-trivial values of $\ptrue$ can be accommodated in the SM.  A SFF with a 10\,\iab\ data-set
would be able to provide sub 1\% measurements of $A_i$ in $\phi K^*$. In addition to this, one can
search for \T-odd \CP\ violating asymmetries in triple products constructed from the angular
distributions~\cite{Datta:2003mj}. It has also been suggested that non-SM effects could be manifest in a number of
other observables~\cite{london}. The measured rates of electroweak penguin-dominated \B\ decays to final states
involving a $\phi$ meson are also probes of NP~\cite{Lu:2006nz}. The study of $B\to AV$ decays (where $A$ is an
axial-vector meson) also provides this rich set of observables to study, however current results only yield an upper
limit on $B^0\to a_1^\pm\rho^\mp$ decays~\cite{babar_a1rho}. \babar\ have recently studied the angular distribution for
$B \to \phi K^*(1430)$~\cite{babar_phikst}.

\section{\boldmath \CPT}

The combined symmetry \CPT\ is conserved in locally gauge invariant quantum field
theory~\cite{luders,pauli,jost,dyson}. It is possible to construct theories of quantum gravity where Lorentz symmetry
breaks down and the quantum coherence of the $\BBbar$ state produced in \Uns{4} decays is broken (for example
see~\cite{kostelecky,mavromatos}). A test of \CPT\ is one of the fundamental tests of nature that should be performed
to increasingly greater precision. Current experimental constraints on \CPT\ in correlated $P^0\overline{P}^0$ system
have been performed for neutral $K$ and $B$ mesons where the most stringent limit on \CPT\ violation from $\B$ decays
is discussed in Ref.~\cite{dileptons}.

\section{CHARM DECAYS}

Several reviews of charm physics~\cite{cicerone} have recently highlighted the motivation to revisit studies of charm
meson decays at much higher luminosities.  The proceedings of the talks within this conference give an overview of the
state of the art measurements in charm decays~\cite{charm_at_beauty2006}.  The motivation for studying charm decays at
a SFF includes the continued search for $D$-mixing and \CPV.  One often neglected fact is that the multitude of
precision charm measurements are instrumental to honing theoretical calculations used in the study of \B\ meson decays.
Charm physics is an integral part of the wider program pursued at a SFF.

\section{\boldmath STUDY OF $\tau$ LEPTONS}

One of the most promising channels to experimentally constrain lepton flavor violation (LFV) in $\tau$ lepton decay is
the process $\tau \to \mu \gamma$.  The current experimental branching ratio limits on this process are
${\cal{O}}(10^{-7})$~\cite{babar_tautomugamma,belle_tautomugamma} using approximately $1.5\times 10^{9}$ $\tau$ pairs.
An estimated $10\times 10^9$ $\tau$ pairs will be produced each year at a SFF.  The large number of recorded decays
would enable one to push experimental sensitivities of LFV down to the $10^{-9}$ to $10^{-10}$ level.  Such a stringent
limit on LFV would impose serious constraints on many models of NP~\cite{cernstrategy}. In addition to LFV, one can
search for \CPV\ in $\tau$ decay.

\section{$\Upsilon$ DECAYS BELOW THE 4S RESONANCE}

Samples of $\Uns{1S,2}$ decays can be obtained by operating a SFF at the $\Uns{3}$ resonance and tagging the final
state $\pi^+\pi^- \Uns{1S,2}$, or via radiative return from the $\Uns{4}$ resonance.

The decays $\Uns{3}\to \pi^+\pi^- \Uns{1S,2}$ with $\Uns{1S,2}\to l^+l^-$ for $l=e, \mu, \tau$, have been proposed for
testing lepton universality (LU) at the percent level using the existing \B-factories~\cite{sanchis_lozarno}.  The CLEO
collaboration have recently performed such measurements for $\tau$ and $\mu$ dilepton decays of $\Uns{1S,2S,3}$
concluding that LU holds within the ${\cal O}(10\%)$ precision of the measurement~\cite{CLEO_UpsilonDILEPTONDECAYS}.
CLEO analysed ${\cal O}(1.1 \ifb)$ of data accumulated at each of the \Uns{1}, \Uns{2} and \Uns{3} resonances. The data
currently show a $2.6\sigma$ deviation from the expectation of LU. Various NP scenarios exist where light \CP-odd
non-SM Higgs bosons could break LU~\cite{sanchis_lozarnob,sanchis_lozarnoc,sanchis_lozarnod}. A precision test of LU
could be performed at a SFF by operating the accelerator at $\sqrt{s}=10.355 \gev$ corresponding to the $\Uns{3}$
resonance.

Most dark matter scenarios require a SM-dark matter coupling, and studies of the decays $\Uns{1}\to
{\mathrm{invisible}}$ have been proposed in order to study such couplings~\cite{mcelrath}.

\section{\boldmath ACCUMULATING DATA AT THE $\Uns{5}$ RESONANCE}

Recent work has shown that it is possible to study $B_s$ decays produced at the \Uns{5} resonance with an asymmetric
energy \epem\ collider~\cite{belle_fives}.  It will be possible to measure $\Delta \Gamma/\Gamma$ for the $B_s$ system
using $B_s \to D_s^{(*)}D_s^{(*)}$ decays~\cite{BNM}, and an \epem\ collider provides a clean environment to
search for NP in the $\Bs\to K^*\gamma$ and $\Bs\to\phi\gamma$ loop processes~\cite{moreBNM}. One can also constrain NP
parameter space through measurements of semi-leptonic \Bs\ decays.  The large mixing frequency of $\Bs$ mesons makes
time-dependent \CP\ asymmetry measurements challenging, and studies are underway to elucidate the prospects of such
measurements with a SFF.

\section{ACCELERATOR DESIGNS}

The \pep2\ and KEK-B asymmetric energy \epem\ accelerators have outperformed expectations to integrate a combined
luminosity of 1\,\iab\ since the \B-factory operation started in 1999.  There are currently two designs being entertained
for a \tentothethirtysix\ collider that would integrate 50\,\iab\ of data during their lifetimes.  One is an upgraded KEK-B
accelerator~\cite{Oide} (Super KEK-B) that benefits from ILC technological developments,
and the other is the result of more recent developments in trying to harness more ILC-related technology~\cite{seeman}
(low emittance design).
Highlights of the proposed parameter sets of these machines are summarised in Table~\ref{tbl:accparams}.
The luminosity ${\cal L}$ of an \epem\ collider is proportional to $I_{e^+/e^-} \xi_{e^+/e^-, y}/\beta^*_y$ where
$I_{e^+/e^-}$ is the beam current, $\xi_{e^+/e^-, y}$ is the beam-beam parameter and $\beta^*_y$ is the vertical
beta-function amplitude at the interaction point.  In addition to this, there is a luminosity reduction factor, the so-called
hourglass effect~\cite{hourglass}, to consider when simulating the delivered ${\cal L}$ from a SFF.  This reduction factor is
 approximately 6\% in the case of PEP-II.
The luminosity increase of the Super KEK-B design is achieved by increasing the beam-beam term, the beam currents and
reducing $\beta^*_y$. In order to reach the predicted luminosity of $0.8\times\tentothethirtysix$, the KEK-based design
incorporates a number of upgrades including the use of crab cavities to rotate the colliding bunches of electrons and
positrons.  This technology is expected to reduce the geometric reduction factor of the luminosity.  The ``low emittance design''
achieves its luminosity increase to $1.0\times\tentothethirtysix$ through a low emittance operation of the accelerator.  While
most of the studies for this design are focussed on a possible site near Frascati, this is a site-independent design.
An important aspect of achieving \tentothethirtysix\ is the use of so called `crabbed waist''
scheme~\cite{crabbedwaist}. 
\begin{table}[!h]
\caption{Parameters of the accelerator configurations under consideration.}\label{tbl:accparams}
\begin{center}
\begin{tabular}{|l|c|c|} \hline
Parameter           & Super KEK-B & low emittance \\ \hline
$\epsilon_x$ (\nm)  & 9.0         & 0.8\\
$\epsilon_y$ (\nm)  & 0.045       & 0.002\\
$\beta^*_x$ (\cm)   & 200.0       & 20.0\\
$\beta^*_y$ (\cm)   & 3.0         & 0.2 \\
$\sigma_z$ (\mm)    & 3.0         & 7.0\\
$I_{e^+}$ (A)       & 9.4         & 2.5 \\
$I_{e^-}$ (A)       & 4.1         & 1.4\\ \hline
\end{tabular}
\end{center}
\end{table}

There are pros and cons for both of the accelerator concepts under study, however as it
is unlikely that there will be more than one SFF built in the world, the next
step for a SFF is to coalesce the best of both designs to a common proposal on a
timescale of the next year or two.  Both designs are able to provide the luminosity required
to achive the physics goals outlined here.  There is more potential for upgrading the low emittance
design to provide higher luminosities than the initial target, however both accelerator designs still
require some R\&D before they can be realised.

\section{SUMMARY}

Both a SFF and an upgraded LHCb experiment can provide an unprecedented precision overconstraint of the
CKM mechanism. While a significant fraction of the physics programs of these two experiments are overlapping,
they also provide a number of complimentary constraints. The $b \to s$ penguin transitions are
golden transitions to search for and constrain NP.  Precision $\Delta S$ measurements of hadronic and radiative penguin
modes such as $B\to\eta^\prime K^0$, $\phi K^0$ and  $B\to K^*\gamma$ will be able to constrain NP loop contributions to
the flavor physics sector. The study of $B\to VV$ decays may shed light on models with right handed couplings, and
electroweak penguin-dominated processes also probe NP in loop processes.  In addition, one can search for
$D^0\overline{D}^0$ mixing, and \CPV\ in charm decays, as well as testing \CPT\ using \B\ mesons.  Dilepton decays of
$\Upsilon(1S)$, \Uns{2}\ and \Uns{3}\ resonances can be used to test LU, and models with dark matter can be tested with
decays to invisible final states. Finally, the unprecedented samples of $\tau$ leptons produced at a SFF provide us
with the opportunity to search for lepton-flavor-violation at sensitivities relevant to most prominent NP scenarios and
also search for \CPV\ in the lepton sector.

It is impossible to predict {\em a priori} what measurements are needed to constrain NP contributions, so one has to
perform as many different measurements as possible.  A SFF would provide a base to perform a wide range of such measurements.

\bibliographystyle{unsrt}
\bibliography{biblio}

\end{document}